\documentclass[aps,prb,showpacs,preprintnumbers,amsmath,amssymb,twocolumn,floatfix,footinbib,superscriptaddress]{revtex4-1}

\usepackage{amsmath,amssymb}
\usepackage{graphicx}
\usepackage{bm}
\usepackage{bbm}
\usepackage{color}
\usepackage{natbib}
\usepackage{hyperref}
\hypersetup{
  colorlinks,
  citecolor=magenta,
  linkcolor=blue,
  urlcolor=blue}

\newcommand{\dd}{\mathrm{d}}
\newcommand{\I}{\mathrm{i}}
\newcommand{\E}{\mathrm{e}}

\newcommand{\cfig}[1]{Fig.~\ref{#1}}
\newcommand{\csec}[1]{Sec.~\ref{#1}}
\newcommand{\ceqn}[1]{Eq.~(\ref{#1})}

\newcommand{\cref}[1]{Ref.~\onlinecite{#1}}
\newcommand{\pd}{\phantom{\dagger}}

\renewcommand{\vec}[1]{\mathbf{#1}}
\renewcommand{\cite}[1]{[\onlinecite{#1}]}

\begin{document}

\title{Quantifying the fragility of unprotected quadratic band crossing points}

\author{Stephan Hesselmann}
\affiliation{Institut f\"ur Theoretische Festk\"orperphysik, JARA-FIT and JARA-HPC, RWTH Aachen University, 52056 Aachen, Germany}
\author{Carsten Honerkamp}
\affiliation{Institut f\"ur Theoretische Festk\"orperphysik, JARA-FIT and JARA-HPC, RWTH Aachen University, 52056 Aachen, Germany}
\author{Stefan Wessel}
\affiliation{Institut f\"ur Theoretische Festk\"orperphysik, JARA-FIT and JARA-HPC, RWTH Aachen University, 52056 Aachen, Germany}
\author{Thomas C. Lang}
\email{thomas.lang@uibk.ac.at}
\affiliation{Institute for Theoretical Physics, University of Innsbruck, 6020 Innsbruck, Austria}

\begin{abstract}
We examine a basic lattice model of interacting fermions that exhibits quadratic band crossing points (QBCPs) in the non-interacting limit. In particular, we consider spinless fermions on the honeycomb lattice with nearest neighbor hopping $t$ and third-nearest neighbor hopping $t''$, which exhibits fine-tuned QBCPs at the corners of the Brillouin zone for ${t'' = t/2}$. In this situation, the density of states remains finite at the Fermi level of the half-filled band and repulsive nearest-neighbor interactions $V$ lead to a charge-density-wave (CDW) instability at infinitesimally small $V$ in the random-phase approximation or mean-field theory. We examine the fragility of the QBCPs against dispersion renormalizations in the ${t\mbox{-}t''\mbox{-}V}$ model using perturbation theory, and find that the $t''$-value needed for the QBCPs increases with $V$ due to the hopping renormalization. However, the instability toward CDW formation always requires a nonzero threshold interaction strength, i.e., one cannot fine-tune $t''$ to recover the QBCPs in the interacting system. These perturbative arguments are supported by quantum Monte Carlo simulations for which we carefully compare the corresponding threshold scales at and beyond the QBCP fine-tuning point. From this analysis, we thus gain a quantitative microscopic understanding of the fragility of the QBCPs in this basic interacting fermion system.  
\end{abstract}

\date{\today}

\maketitle
 
\section{Introduction}

The vast majority of itinerant fermion systems such as normal metals, or sufficiently doped semiconductors, exhibit extended Fermi surfaces. Compared to the non-interacting reference limit interactions may deform such Fermi surfaces, yet drastic, qualitative or topological changes in the Fermi surface are rare and usually require strong interactions. The situation is remarkably different for systems with small Fermi surfaces, in particular for point-like Fermi surfaces in two dimensions. Unless protected by the point group symmetries of the lattice even weak interactions can split up band crossing points and lead to qualitatively different low-energy behavior, even without the breaking of the underlying symmetries. For a QBCP to be stable without fine tuning the system must be time-reversal invariant and the QBCP must have $C_4$ or $C_6$ rotational symmetry \cite{Sun09}. QBCPs on threefold symmetric lattices such the honeycomb and relatives are however unprotected and prone to such topological Lifshitz transitions. In the (particular) focus of current research are QBCPs on various tailored two-dimensional systems, such as single layer graphene~\cite{Montambaux12}, Bernal-stacked bilayer graphene~\cite{Lemonik10,Scherer12,McCann13} and twisted bilayer graphene~\cite{Santos07,Bistritzer11,Hejazi19}, or more generally as playground for unconventional ordering instabilities~\cite{Uebelacker11,Murray14,Pawlak15,Zheng18}. In fact, basic models for such systems exhibit quadratic band crossing points (QBCPs) at high-symmetry points in the Brillouin zone (BZ). The introduction of interactions may split these into one central and three satellite Dirac points in symmetric directions away from the original QBCP. 

It was recently found~\cite{Pujari16}, that in contrast to the previous assessment~\cite{Lang12}, the rapid loss of antiferromagnetic (AF) order in the Bernal-stacked bilayer honeycomb Hubbard model for weaker interaction strengths in quantum Monte Carlo (QMC) simulations may be caused by the split of the QBCPs from the non-interacting limit. A perturbative calculation of the self-energy effects for the lattice model~\cite{Honerkamp17} as well as a renormalization group study for the continuum model~\cite{Ray18} confirmed the mechanism for the interaction-driven dispersion deformation. Yet, the lattice study~\cite{Honerkamp17} suggested a much lower threshold interaction strength for the destruction of the ordered ground state than the numerical observations by QMC, due to rather weak self-energy effects in the bilayer honeycomb Hubbard model.

The reason for the diminished self-energy effects that change the dispersion is that, for the Hubbard model, they only occur in second order in the on-site interaction strength $U$. This leads to a low Lifshitz energy scale $\epsilon_L$ for the dispersion renormalization, which is quadratic in the relevant hopping renormalization and hence scales only ${\sim U^4}$. In contrast to this, the AF ground state order is an effect that is driven by $U$ already in first order and is hence numerically dominant. The loss of AF order upon decreasing $U$ is roughly determined by the crossing between the Lifshitz energy ${\epsilon_L \sim U^4}$ and the energy scale ${\sim \E^{-1/(U\rho^*)}}$ of the AF order (here, $\rho^*$ is an effective density of states) \cite{Honerkamp17}. For an interlayer hopping of the order of $t$, as used in \cref{Pujari16} this crossing resides at a small value of ${U \lesssim t}$, involving low energy scales ${\sim 10^{-6}t}$. It thus will be hard to see this effect in a numerical simulation on a finite lattice, or to associate any stronger observable feature with it.

As we will show in the following, this situation improves if we instead consider the single-layer honeycomb ${t\mbox{-}t''\mbox{-}V}$ model. This system exhibits QBCPs at both Dirac points, $K$ and $K'$, of the BZ, when the hopping between third-nearest neighbors ${t'' = t/2}$ takes half of the value of the nearest-neighbor hopping $t$ \cite{Montambaux12}. In this case, already the first-order Fock self-energy causes a renormalization of the nearest-neighbor hopping ${t \to t + \delta t}$, with ${\delta t \approx 0.2 V}$. This effect disturbs the fine-tuning of ${t'' = t/2}$ and destroys the original QBCPs. Moreover, the energy scale below which the density of states drops to zero linearly is found to scale as ${\epsilon_L/t \sim \left[ t''/(t+\delta t) - {1}/{2}\right]^2}$, i.e. quadratic in $V$. Therefore, it can be a numerically much larger scale. This positions the onset of charge density wave (CDW) order due to the nearest-neighbor interaction towards higher threshold values $V_c$.  

The emerging relevant energy scales in this specific system enable us, for the first time, to perform reliable QMC simulations and to provide a consistent statement on the nature of the fragility of unprotected QBCPs from perturbation theory and numerical simulations. In the remainder of this paper, we quantify this scenario using low-order perturbation theory and also provide supportive numerical evidence from unbiased QMC simulations.

In the following \csec{model} we introduce the spinless fermion ${t\mbox{-}t''\mbox{-}V}$ model on the honeycomb lattice and discuss the potential charge density wave (CDW) instability at the random phase approximation (RPA) level in \csec{cdw}, accounting for the lowest-order self-energy corrections in \csec{fock} and perform self-consistent first-order perturbation theory in \csec{pert}. Section \ref{qmc}  provides supporting evidence to the perturbative arguments from quantum Monte Carlo calculations and analyzes finite size effects in comparison with mean field results.

\section{The ${t\mbox{-}t''\mbox{-}V}$ Model and its QBCPs \label{model}} 

\begin{figure}
	\centering
	\includegraphics[width=\columnwidth]{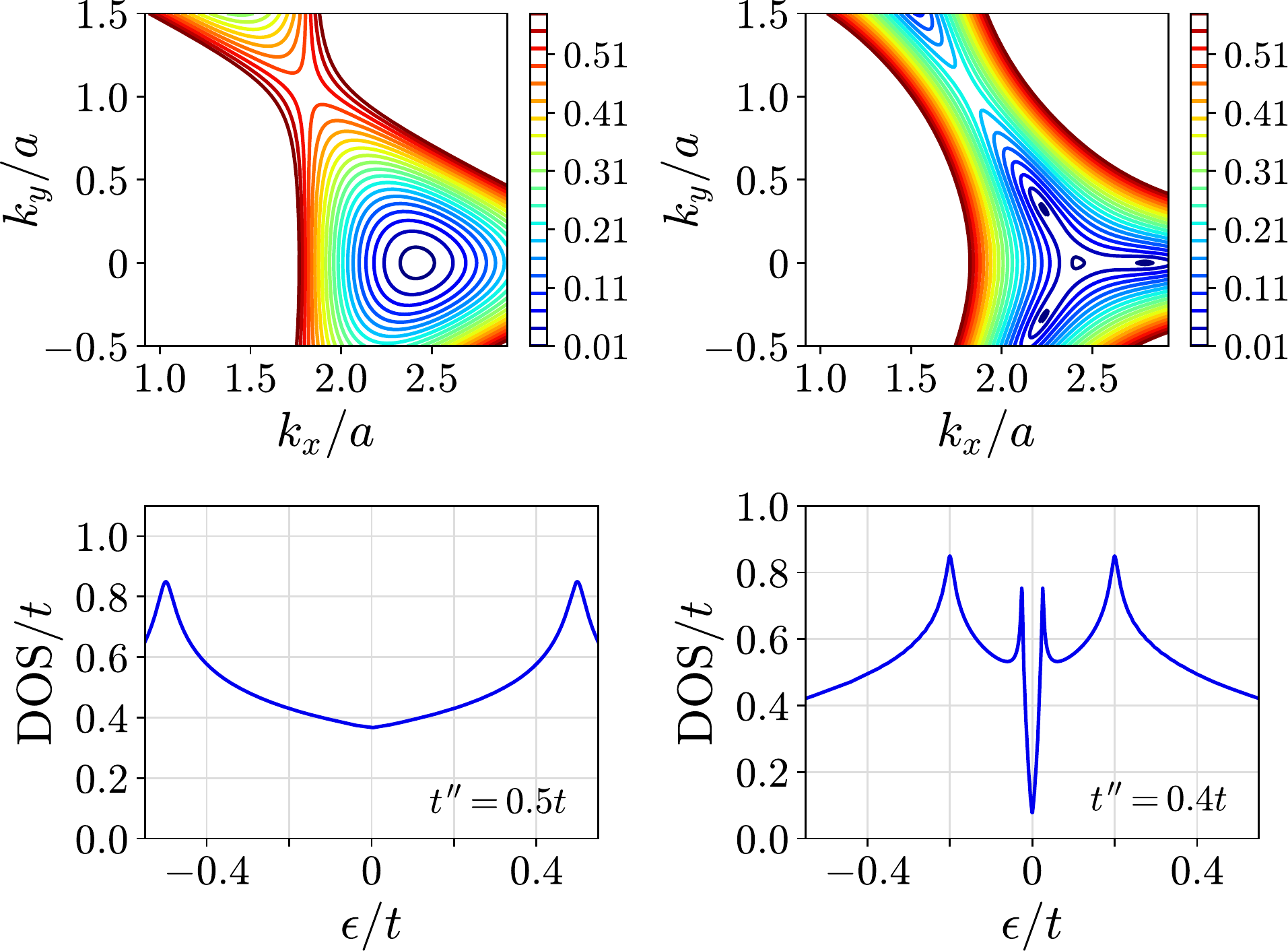}%
	\caption{Contour plots of lines of constant positive band energy (top panels) near the Dirac point at $K\approx (2.42 ,0)$ and the corresponding DOS (bottom panels) for the non-interacting model, at ${t'' = 0.4t}$ (left panels) and ${t'' = 0.5t}$ (right panels). In the left lower panel, the DOS drops to zero linearly below ${\epsilon \approx 0.025t}$. It does not reach zero because of the finite numerical resolution. The right lower plot shows the DOS for the QBCP condition. \label{dispdos}}
\end{figure}

We consider spinless fermions on the honeycomb lattice, with a hopping kinetic energy
\begin{equation}\label{H0}
   H_0 = - t \sum_{\langle ij \rangle} c_i^{\dagger} c_j^{\pd} - t'' \sum_{\langle\!\langle ij \rangle\!\rangle} \left( c_i^{\dagger} c_j^{\pd} + \mathrm{H.c.}\right) \;,
\end{equation}
where $t$ is the hopping amplitude between nearest-neighbor sites on the lattice, which belong to the two different sublattices $A$, $B$, and the hopping amplitudes $t''$ is between third-nearest neighbors on the honeycomb lattice, which are second-nearest neighbors within the same sublattice, straight across the hexagons of the honeycomb lattice. We note that the vanishing second-nearest neighbor hopping $t'=0$ ensures the bipartite lattice structure, and is protected by particle-hole symmetry. With the Bravais lattice vectors ${\vec{a}_1 = (\sqrt{3}, 0)}$ and ${\vec{a}_2 = (\sqrt{3}/2, 3/2)}$ translated into momentum space these two hopping terms give rise to the ${2 \times 2}$ matrix Hamilton
\begin{equation}\label{Hk0}
   H_0 = \sum_{\vec{k}} \left(\!\begin{array}{c} a_{\vec{k}} \\ b_{\vec{k}} \end{array}\!\right)^\dagger \cdot
       \left(\!\begin{array}{cc} 0 & h(\vec{k}) \\ h^* (\vec{k}) & 0 \end{array}\!\right) \cdot
       \left(\!\begin{array}{c} a_{\vec{k}} \\ b_{\vec{k}} \end{array}\!\right) \;, 
\end{equation}
with ${h(\vec{k}) = t\;\!h_1(\vec{k} ) + t''\, h_2(\vec{k})}$, where
\begin{eqnarray}
   h_1(\vec{k}) &=& 1 + 2 \cos \left( \frac{\sqrt{3}}{2} k_x \right) \, \E^{\I \frac{3}{2} k_y} \;,\;\mbox{and}\label{h1k}\\
   h_2(\vec{k}) &=& \E^{\I 3 k_y} + 2 \cos \left( \sqrt{3} k_x \right) \;.\label{h2k}
\end{eqnarray}
In this setup, with the lattice constant set to unity, the Dirac points are located at ${K = \left(2\pi/3, 2\pi/\sqrt{3}\right)/\sqrt{3}}$ and ${K' = \left(4\pi/3,0\right)/\sqrt{3}}$. Expanded around these points for small momenta $\vec{p}$, both hopping form factors $h_1(\vec{k})$ and $h_2(\vec{k})$ grow linearly ${\propto \pm\, p_x + \I p_y}$. For ${t'' = t/2}$, these two contributions cancel exactly. In this situation, the remaining quadratic terms in $p_{x/y}$ dominate near $K$ and $K'$, leading to QBCPs. The corresponding dispersions and densities of states (DOS) at low energies for ${t'' = 0.5t}$ and ${t'' = 0.4t}$ are shown in \cfig{dispdos}. In this paper, we restrict our study to half band filling, i.e., the chemical potential is zero.

As an interaction, we consider the nearest-neighbor repulsion $V$
\begin{equation}\label{HV}
   H_V = V \sum_{\langle ij \rangle} c_i^{\dagger} c_j^{\dagger} c_j^{\pd} c_i^{\pd} = \frac{1}{N} \sum_{\vec{q}} V(\vec{q})\, n^A_{\vec{q}}\,n^B_{-\vec{q} } \;,
\end{equation} 
with sublattice resolved density operators $n^{A/B}_{\vec{q}}$. Here, $N$ is the number of unit cells and the form factor $V(\vec{q})$ is
\begin{equation}\label{Vq}
   V(\vec{q}) = V \left[ 1+ 2 \cos \left( \frac{\sqrt{3}}{2} q_x \right) \, \E^{\I \frac{3}{2} q_y} \right] \,,
\end{equation}
with ${V(\vec{q}) \rightarrow 3V}$ for ${\vec{q} \to 0}$ in the long-wavelength limit.

A detuning of $t''$ from the value of ${t''=t/2}$ causes the QBCPs to split up into four Dirac points. In the detuned case (\cfig{dispdos}, right panels), the dispersion is found to go through a maximum between the central Dirac point and each of the outer ones at an energy ${\epsilon_L \sim c (t''/t-1/2)^2}$, where the prefactor ${c\approx 2t}$. This is also the energy scale at which a peak resides in the density of states, and below which the dispersion becomes linear and the DOS therefore decreases linearly to zero. 

\section{CDW instability \label{cdw}}

\begin{figure}
	\centering
	\includegraphics[width=\columnwidth]{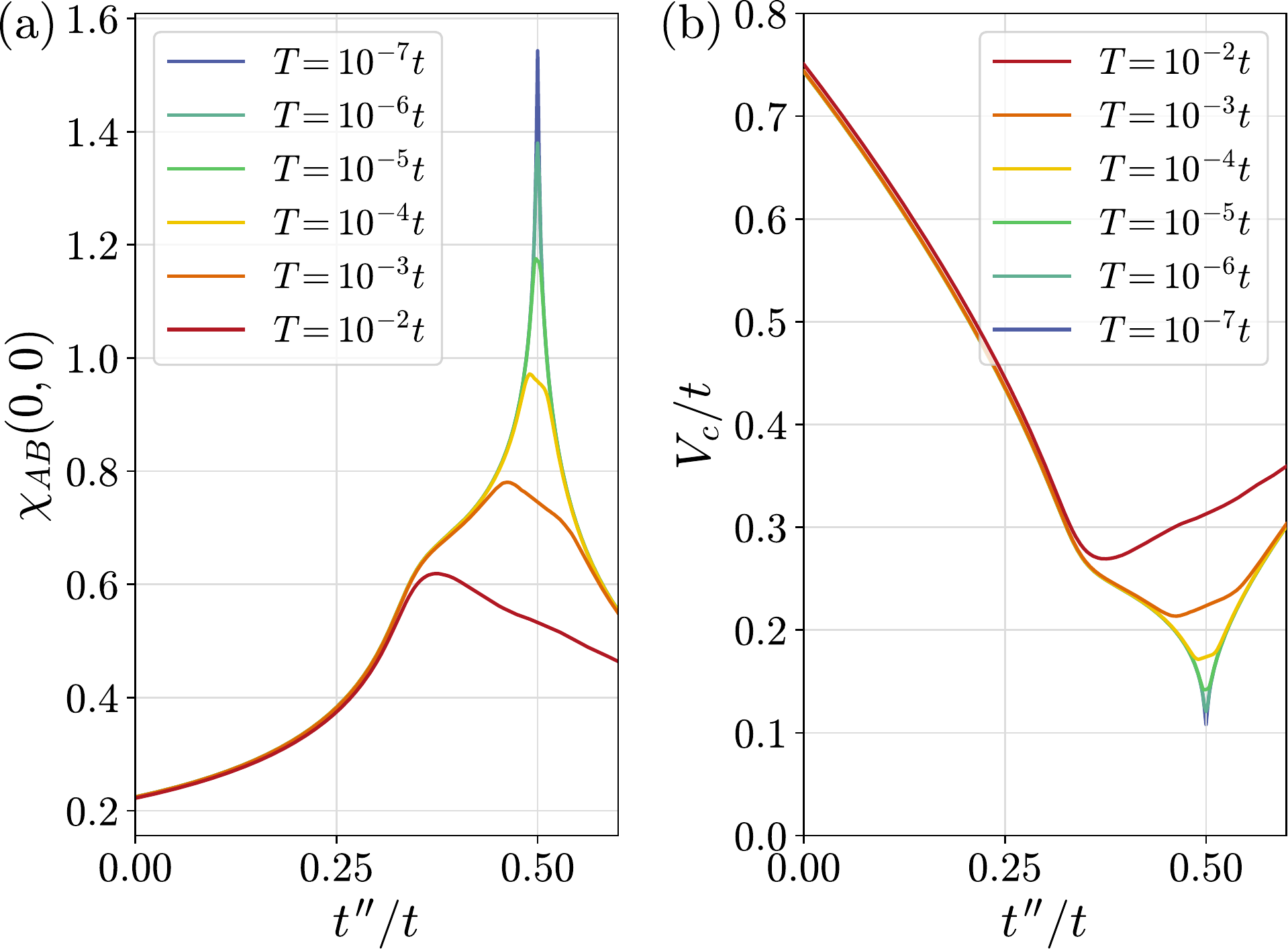}%
	\caption{(a) Static non-interacting CDW susceptibility as a function of $t''$ for different temperatures. (b) The corresponding results for the threshold value $V_c$, according to the Stoner criterion.
	\label{chiVcplot}}
\end{figure}

The nearest-neighbor repulsion $V$ eventually induces CDW order in the half-filled system. We first consider the susceptibility of the free case towards CDW order. For this purpose, we calculate the ${2\times 2}$ bare susceptibility matrix ${\chi (\vec{q},\nu)}$ for the different sublattice combinations with ${o,o' \in \{A,B\}}$,
\begin{equation}\label{chi0oo'}
   \chi^{(0)}_{oo'}(\vec{q},\nu) = \int_0^\beta \, \dd\tau \, \E^{\I \nu \tau} \langle n^o_{\vec{q}} (\tau)\, n^{o'}_{-\vec{q}} (0) \rangle \;. 
\end{equation}
In the random phase approximation (RPA) for the interaction introduced in \ceqn{Vq}, the interacting susceptibility becomes 
\begin{equation}\label{chiRPA}
   \chi (\vec{q},\nu) = \chi^{(0)} (\vec{q},\nu) \left[ \mathbbm{1} - V(\vec{q}) \chi^{(0)} (\vec{q},\nu)\right]^{-1} \;,
\end{equation}
with the interaction matrix 
\begin{equation}\label{vqmatrix}
   V(\vec{q}) = \left(\! \begin{array}{cc} 0 & V(q) \\ V^*(q) & 0 \end{array} \!\right) \;. 
\end{equation}
The CDW order occurs at $\vec{q}=0$ and $\nu=0$. For systems with the free Hamiltonian as in \ceqn{Hk0}, one can show that
\begin{equation}\label{chiaachiab}
   \chi^{(0)}_{AB}(0,0) = \chi^{(0)}_{BA}(0,0) = - \chi^{(0)}_{AA}(0,0) = - \chi^{(0)}_{BB}(0,0) \;.
\end{equation}
Then the inverse of the matrix $D = \left[ \mathbbm{1} - V(\vec{q})\, \chi^{(0)} (\vec{q},\nu)\right]$ in \ceqn{chiRPA} becomes singular when the Stoner criterion
\begin{equation}\label{stoner}
   \frac{1}{V} = 6\, \chi^{(0)}_{AB}(0,0) \;,
\end{equation}
is satisfied. The eigenvalue of $D$ that approaches zero for $V \to V_c$ belongs to an eigenvector that changes sign between the two sublattices. This clearly identifies the CDW as the emerging ordering tendency, since the susceptibility with respect to a staggered density field becomes singular. The factor of 6 in \ceqn{stoner} is a consequence of the two sublattices (providing a factor of 3) and the interaction in \ceqn{Vq} for ${\vec q \to 0}$ (providing the other factor of 2). At ${t'' = 0}$, the susceptibility approaches ${\chi^{(0)}_{AB}(0,0) \approx 0.226/t}$, which results in the  RPA threshold value ${V_{c,0} \approx 0.74t}$ for the CDW ordering transition. In comparison, for  the spinfull Hubbard model, with an on-site interaction $U$ and thus without a factor of 3 from the nearest neighbors, the threshold interaction strength in  RPA for the AF order is ${U_c \approx 2.21t}$, in accordance with earlier literature \cite{Sorella12}. In fact, the above RPA results for the critical interaction strength agree exactly with the results from Hartree-Fock mean-field theory for the onset of the corresponding ordering. Quantum corrections that extend beyond mean-field theory (i.e., channel coupling, dispersion renormalization and quasiparticle degradation) increase these threshold values, to ${V_c \approx 1.36t}$ for CDW order in the spinless fermion $t$-$V$ model \cite{Wang14,Li15,Hesselmann16} and to ${U_c \approx 3.8t}$ for the AF transition in the Hubbard model \cite{Meng10,Sorella12}, as reported from QMC simulations. 

The   susceptibility ${\chi^{(0)}_{AB}(0,0)}$ can be computed numerically for all values of $t''$. For the QBCP condition at ${t'' = t/2}$, it diverges logarithmically at low $T$, due to the perfect nesting (for $\vec q =0$) and the finite density of states. Data for different temperatures $T$ are shown in \cfig{chiVcplot}(a). One can clearly observe a sharpening peak at ${t'' = t/2}$, but one also sees that rather low values of $T$ are needed in order to resolve the singularity of the bubble. Figure~\ref{chiVcplot}(b) shows the critical interaction strength $V_c$ as computed from the Stoner criterion in \ceqn{stoner}. At low $T$ and for $t'' \to t/2$, the threshold value $V_c$ dives toward zero to eventually become infinitesimally small for perfect fine-tuning towards the QBCP condition at ${t'' = t/2}$.

\begin{figure}
	\centering
	\includegraphics[width=\columnwidth]{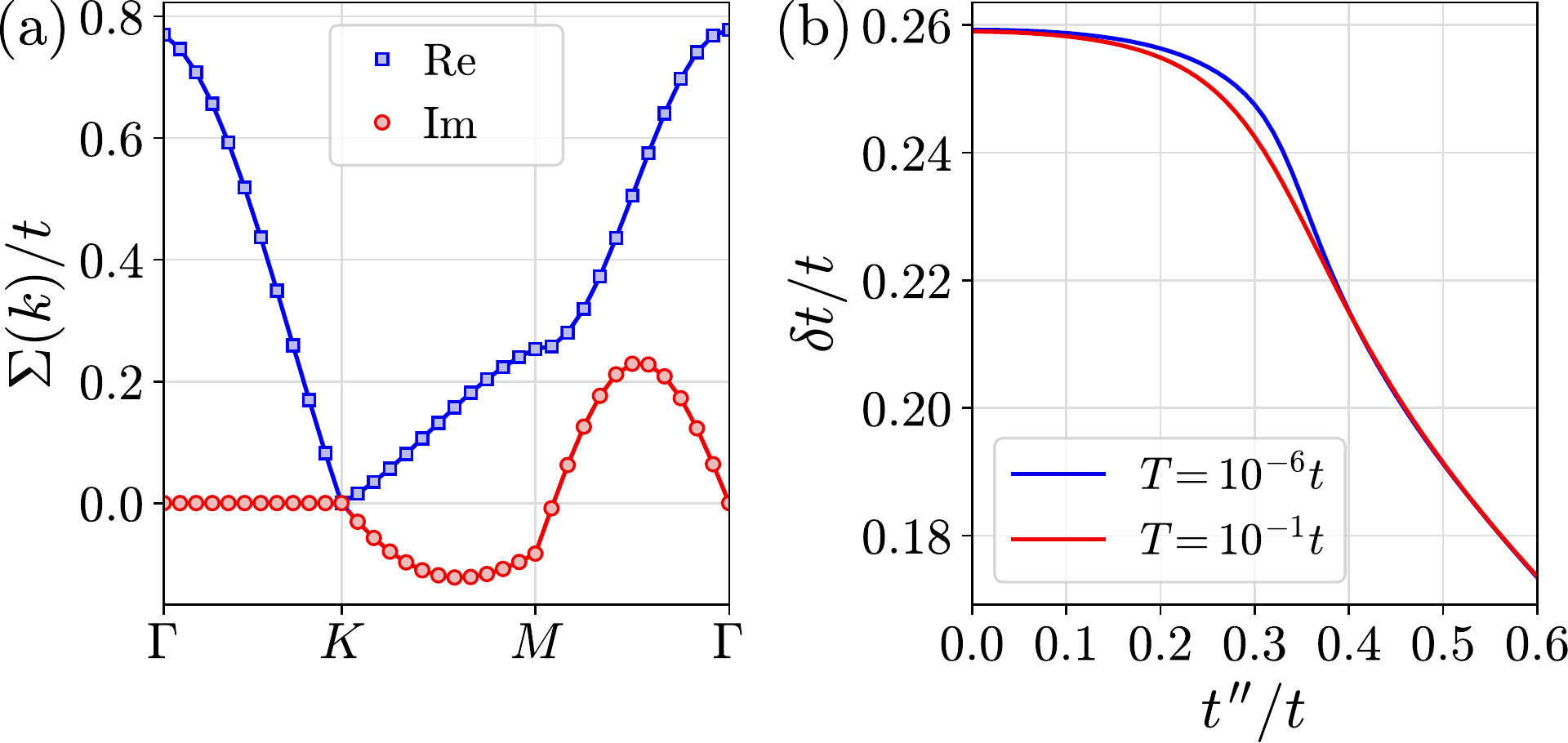}%
	\caption{(a) Real and imaginary part of the (non-self-consistent) momentum-dependent Fock self-energy along the edge of the irreducible Brillouin zone for ${V = t}$. The underlying lines show the real and imaginary parts of the nearest-neighbor hopping form-factor $h_1(\vec{k})$ in \ceqn{h1k}. (b) Hopping renormalization ${\delta t}$ from the first-order self-energy contribution for ${V = t}$, calculated using different values of $T$.
	\label{sigma1storder}}
\end{figure}

\section{Fock self-energy \label{fock}}

The lowest-order self-energy of the spinless fermion model is given by the Fock diagram. For general interactions of the form in \ceqn{HV} it is frequency-independent and reads
\begin{equation}\label{key}
   \Sigma_{oo'} (\vec{k}) =- \frac{T}{N} \sum_{\vec{k}',i\omega'} V_{oo'} (\vec{k}-\vec{k}')\, G_{oo'} (\vec{k}',\I \omega' ) \;. 
\end{equation}
The Matsubara sum can be performed analytically, and since the interaction is only between different sublattices, we only need to evaluate
\begin{equation}\label{focksigma}
   \Sigma_{AB} (\vec{k}) =- \frac{1}{N} \sum_{\vec{k}',\ell} V(\vec{k}'-\vec{k}) \, u_{A\ell}(\vec{k}')\, u_{B\ell}(\vec{k}') \, n_F\left[\epsilon_{\ell}(\vec{k}')\right] \;,
\end{equation}
where the sum extends over the two bands ${\ell = 1,2}$. The numerical evaluation of this expression gives a $\vec{k}$-dependence that almost perfectly follows the form factor $h_1(\vec{k})$ of the nearest-neighbor hopping in \ceqn{h1k} for all $t''$, as shown in \cfig{sigma1storder}(a). This is not surprising, as the self-energy expressed in real space merely collects the Green's functions to nearest neighbors, multiplied by the value of the corresponding interaction strength. Hence, we can straightforwardly read off a nearest-neighbor hopping renormalization ${t\rightarrow t + \delta t}$.  We note that in accord with the particle-hole symmetry of $H_0$, no
 second-nearest neighbor hopping contribution is generated in our perturbation calculation. 

The sign of the correction ${\delta t}$ is such as to enhance the bare hopping amplitude, i.e., the Dirac cones become more acute-angled as one accounts for self-energy corrections. The same trend is of course well known from the long-ranged Coulomb interaction where the renormalization of the velocity $v_k$ becomes very strong near the Dirac point \cite{Gonzalez94,Sharma16}. Here, with short-range interactions, we obtain a \textit{global} finite renormalization of $t$.

In \cfig{sigma1storder}(b) we show ${\delta t}$ as a function of $t''$ for an interaction strength of $V = t$ (due to the first-order nature of this correction, $\delta t$ just scales linearly with $V$). The effect is most pronounced at ${t'' = 0}$, and the corresponding increase of the velocity by about $26\%$ compares remarkably well to the Fermi velocity renormalization reported in \cref{Schuler19} in the low-$V$ region of the $t$-$V$ model from QMC simulations. There is a reduction of ${\delta t}$ in \cfig{sigma1storder}(b) for the larger values of $t''$, but quite roughly, we find ${\delta t \sim 0.2 V}$ for all relevant parameters. In \cref{Schuler19}, the velocity renormalization for the $t$-$V$ model was found to increase to about $35\%$ near the critical interaction strength $V_c$, and thus the above first-order result should be considered a lower bound for the hopping renormalization. 

Note that a similar velocity increase ${\delta v > 0}$ is found from the second- and higher-order self-energy in the spinful Hubbard model on the honeycomb lattice, where the first-order correction to the hopping is zero (cf. \cref{Honerkamp17}). In this case, the renormalized velocity is not necessarily increased. Because of the higher-oder nature of the hopping renormalization in the spinful Hubbard case, the quasiparticle renormalization expressed by a $Z$-factor ${\le 1}$ becomes of the same order and the renormalized velocity ${Z(v_k+\delta v)}$ may actually come out smaller than the bare value $v_k$.

\section{Self-consistent first-order perturbation theory \label{pert}}

\begin{figure}
	\centering
	\includegraphics[width=\columnwidth]{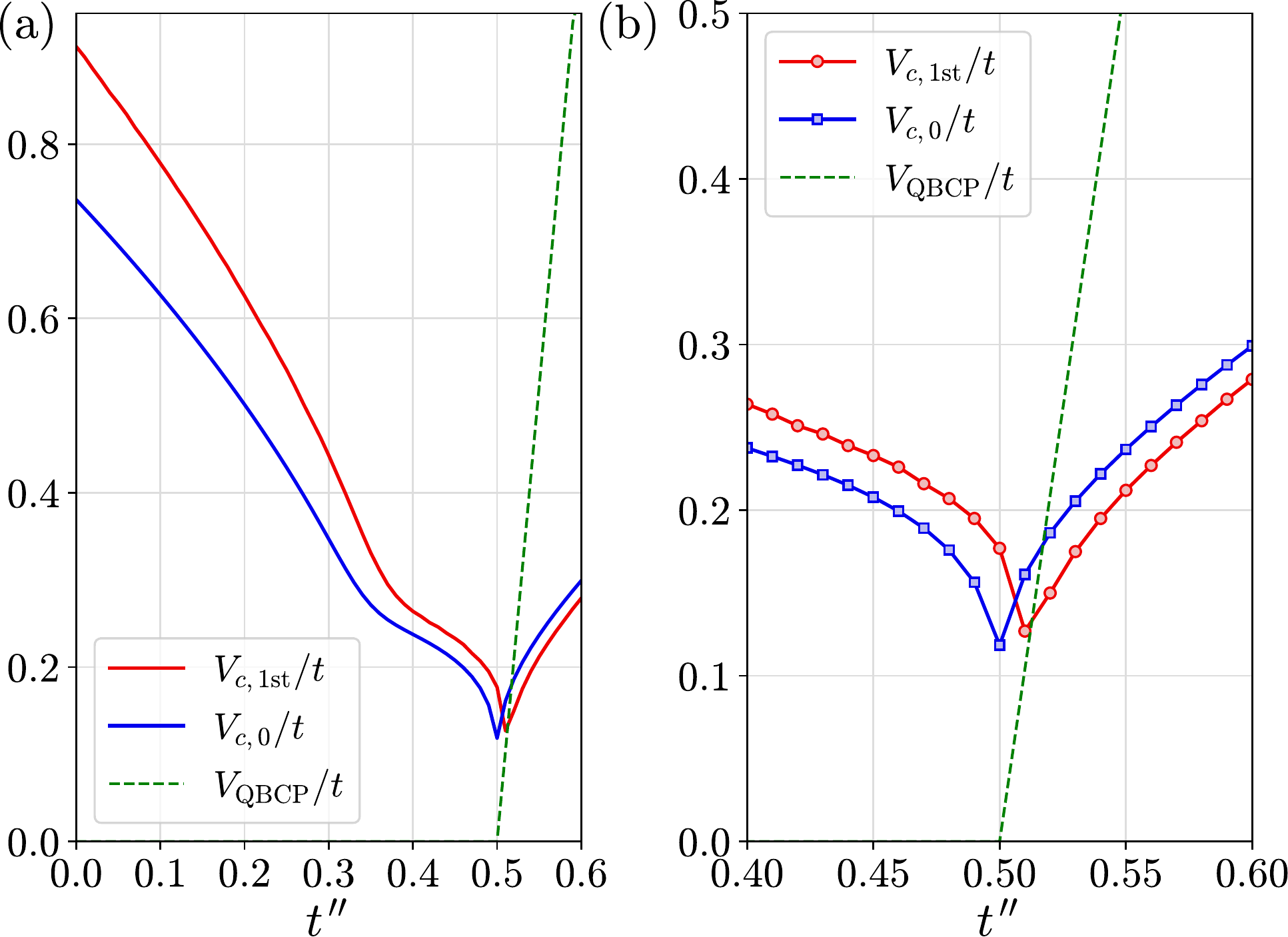}%
	\caption{(a) Comparison of the bare ($V_{c,0}$) and hopping-renormalized (using self-consistent first-order perturbation theory, $V_{c,\mathrm{1st}}$) critical interaction strengths as functions of $t''$, for ${T = 10^{-5}t}$. Also indicated by the dashed line is the position of the QBCP condition ($V_{\mathrm{QBCP}}$) within self-consistent first-order perturbation theory. The panel (b) focuses on the region near ${t'' = 0.5t}$.
	\label{Vc1storder}}
\end{figure}

\begin{figure}
	\centering
	\includegraphics[width=\columnwidth]{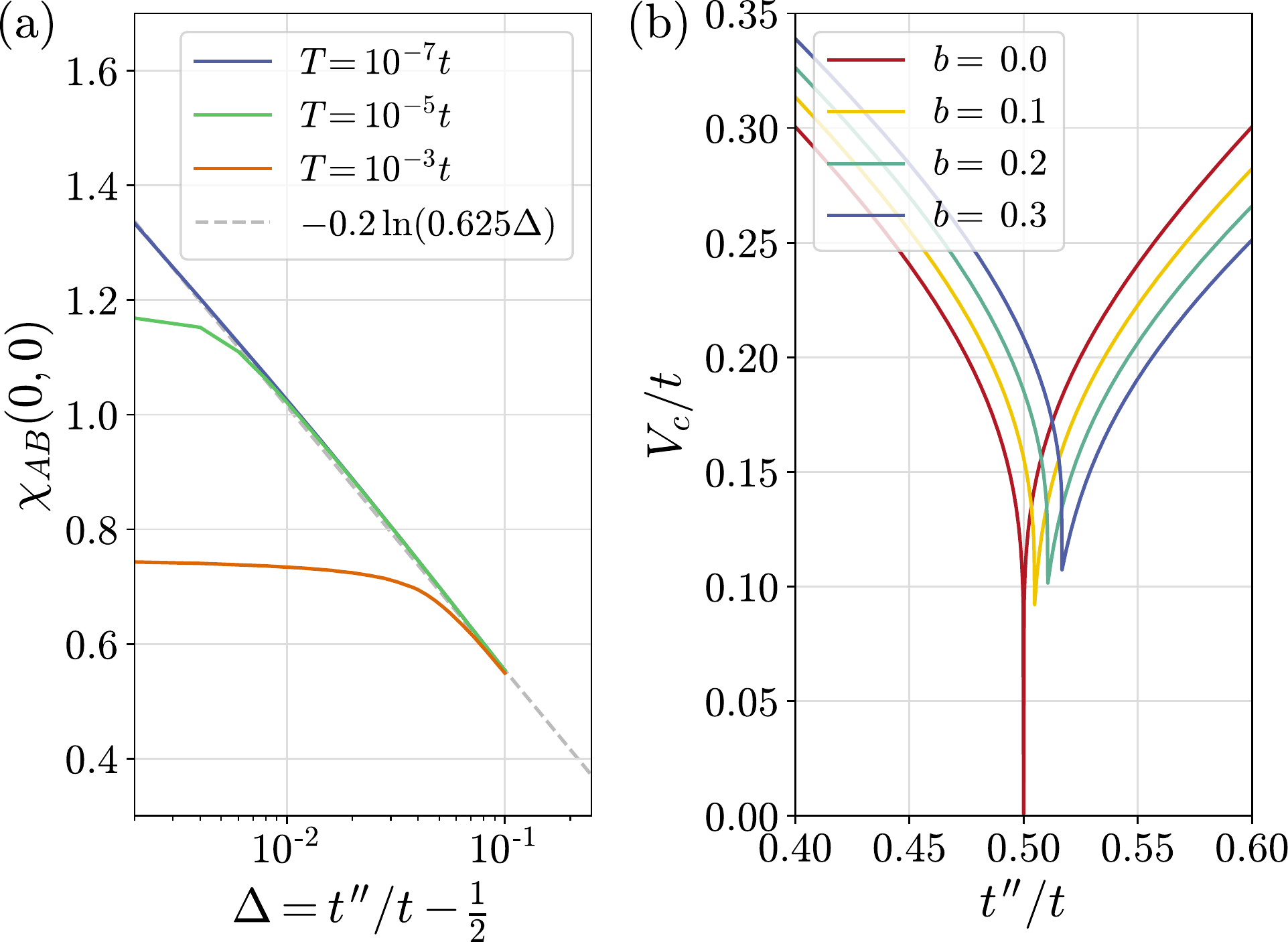}%
	\caption{(a) The static CDW susceptibility as a function of the detuning ${\Delta = t''/t - 1/2}$. Also indicated is a fit of the low-$T$ data to a logarithmic dependence. (b) The Critical interaction values $V_c$ according to the generalized Stoner criterium for different, fixed values of $b$.
	\label{vana}}
\end{figure}

The next step is to perform self-consistent first-order perturbation theory. At ${t'' = 0}$, one finds from \ceqn{focksigma}, using the fact that the matrix elements $u_{A\ell}$ and $u_{B\ell}$ at $t''=0$ are the same for any ${t \not= 0}$, that the self-energy is independent of ${\delta t}$. Hence, the first-order result is already self-consistent in this case. For ${t'' \not= 0}$, we obtain self-consistency either by iteration or by computing the output-${\delta t}$ from the self-energy as a function of an input-${\delta t}$ inserted into the Hamiltonian and requiring equality between the input-${\delta t}$ and the output-${\delta t}$. The results are not far away from and slightly larger than the bare (non-self-consistent) values. In all cases studied for ${0 \le t'' \le 0.6t}$, ${\delta t}$ was found to be positive, i.e., the Dirac cone steepness is enhanced by the self-energy.

The self-energy-related shift ${\delta t}$ disturbs the fine-tuning that is needed to obtain the QBCPs. Without self-energy corrections, the QBCPs were found for ${t'' = t/2}$. However, interactions renormalize $t$ to ${t + \delta t(V)}$. The QBCP is then shifted to occur for ${t_{\mathrm{QBCP}}''(V) = \left[ t+\delta t(V) \right]/2}$, i.e., at values of ${t'' > t/2}$. In \cfig{Vc1storder}, we plot a line $V_{\mathrm{QBCP}}$ as a function of $t''$, which is defined as the interaction strength required to obtain a QBCP in the renormalized bands for a given value of $t''$. We find that $V_{\mathrm{QBCP}}$ rises approximately linear from the non-interacting (${V = 0}$) QBCP case, ${t'' = t/2}$, in quantitative agreement with the mentioned self-energy strength ${\delta t \approx 0.2V}$. Hence, for any weak, nonzero interaction $V$, the renormalized band structure maintains two QBCPs at $K$ and $K'$, but located at somewhat higher and $V$-dependent values of ${t'' > t/2}$.

This leads to the interesting question whether, as QBCPs are found for any $V$ in this regime, there is still a CDW instability at infinitesimally small $V$. The data in Fig.~\ref{Vc1storder} for rather low $T=10^{-6}t$ suggests that this is not the case. Here, we compute the CDW susceptibility $\chi_{AB}(0,0)$ using the renormalized dispersion and determine the threshold interaction strength $V_c$ from the Stoner criterion ${V_c^{-1} = 6 \chi_{AB}(0,0)}$. Compared to the curve for the critical $V_{c,0}$ based on ${\chi^{(0)}_{AB}(0,0)}$, i.e.,  without self-energy corrections, the curve for the first-order ${\delta t}$-corrected $V_{c,\mathrm{1st}}$ is shifted to larger values of $t''$. This in accordance with the arguments above that the QBCPs shifts to larger $t''$ when the hopping correction ${\delta t}$ is taken into account. Of course, one could ask if the dip in $V_{c,\mathrm{1st}}$ at ${t''\approx 0.51t}$ would actually go to ${V = 0}$ for even lower ${T \to 0}$. However, this cannot happen, since for ${V \to 0}$ there is no quadratic dispersion at ${t'' \not=t/2}$, and the logarithmic divergence of the CDW susceptibility is thus cut off.

Note that for most of the range ${t'' > t/2}$, the value for $V_{c,\mathrm{1st}}$ lies below $V_{\mathrm{QBCP}}$. Only very close to the minimal value, is a slightly larger interaction than $V_{\mathrm{QBCP}}$ required, i.e., the situation with a DOS at the Fermi level is still stable. This numerical finding is most likely due to the nonzero temperature used in the computation of the susceptibility. 

The situation can be examined analytically based on a generalized Stoner criterion, which takes into account the 
interaction-generated $\delta t$. The main ingredient for this analysis is the CDW susceptibility and how it depends on the detuning ${\Delta = t''/t - 1/2}$ from the QBCP situation. The self-energy effect can then be included by ${t \to t + \delta t}$. As shown in the left part of Fig. \ref{vana}, at low temperatures the susceptibility $\chi_{AB}(0,0)$ depends logarithmically on the detuning, 
\begin{equation}\label{chifit}
   \chi_{AB}(0,0) \approx - 0.2 \ln \left| 0.625\left( \frac{t''}{t} - \frac{1}{2} \right) \right| \;. 
\end{equation}
This qualitative behavior is obvious, as the density of states starts to vanish linearly below the Lifshitz energy scale ${\epsilon_L\sim (t''/t - \frac{1}{2})^2}$, which acts as a cutoff in the logarithm. The energies with a linear DOS below this scale provide a finite contribution to the susceptibility. With the hopping renormalization ${\delta t = b V}$ in first order perturbation theory (where ${b \approx 0.2}$) and replacing ${t \to t + \delta t}$ in \ceqn{chifit}, we arrive at the generalized (renormalized) Stoner criterion for the CDW instability,
\begin{equation}\label{renstoner}
   \frac{1}{V} = -0.2 \ln \left| 0.625\left( \frac{t''}{t+bV} - \frac{1}{2} \right) \right| \;.
\end{equation} 
The qualitative behavior of \ceqn{renstoner} can be found by considering a plot of the left and the right hand side as functions of $V$ for a given value of $t''$. For large values of $V$, the left hand side vanishes, while the right hand side approaches a nonzero constant provided that ${b > 0}$, which we indeed found above. Therefore, the right hand side resides above the left hand side for large values of $V$. Let us now consider the behavior of both sides of \ceqn{renstoner} upon lowering $V$. The logarithm on the right hand side rises more slowly than any negative power for small arguments and hence than $1/V$. Thus, eventually, upon decreasing ${V \to 0}$, the steeply rising left hand side cuts through from below. Its crossing point with the right hand side defines the critical interaction strength $V_c$ for the considered value of $t''$. As becomes clear from this argument, for any ${b > 0}$, $V_c$ is always nonzero. It is smaller for ${t'' > t/2}$, since then the logarithm on the right hand side can still diverge and the crossing point is reached only at smaller values of $V$. For ${t''< t/2}$, on the other hand, the logarithm remains finite, and the crossing point is located at a larger value of $V$. A crossing point at ${V \to 0}$ requires the argument of the logarithm to go to zero exponentially in $V$. This is not the case in any finite order of perturbation theory for ${\delta t}$. Even upon approaching the CDW instability, the infinite-order effective interaction only diverges like a power law. In principle, for ${b = 0}$ there would still be an isolated solution of \ceqn{renstoner} at exactly ${V = 0}$, but this is physically irrelevant for any interacting system. Hence, for any ${V > 0}$ and $t''$ in the model considered, we can rule out a CDW instability at infinitesimally small interaction strength.

The numerical solution of \ceqn{renstoner} is shown in \cfig{vana}(b), considering three different, fixed values of $b$. This plot confirms the qualitative arguments given above: The curve for ${b = 0}$, i.e., without self-energy corrections, approaches ${V_c=0}$ at ${t'' = t/2}$, while all curves with a finite value of ${b > 0}$ exhibit a nonzero minimal interaction strength for the instability to occur according to \ceqn{renstoner}.

\section{QMC analysis \label{qmc}}

\begin{figure}[t]
	\centering
	\includegraphics[width=0.96\columnwidth]{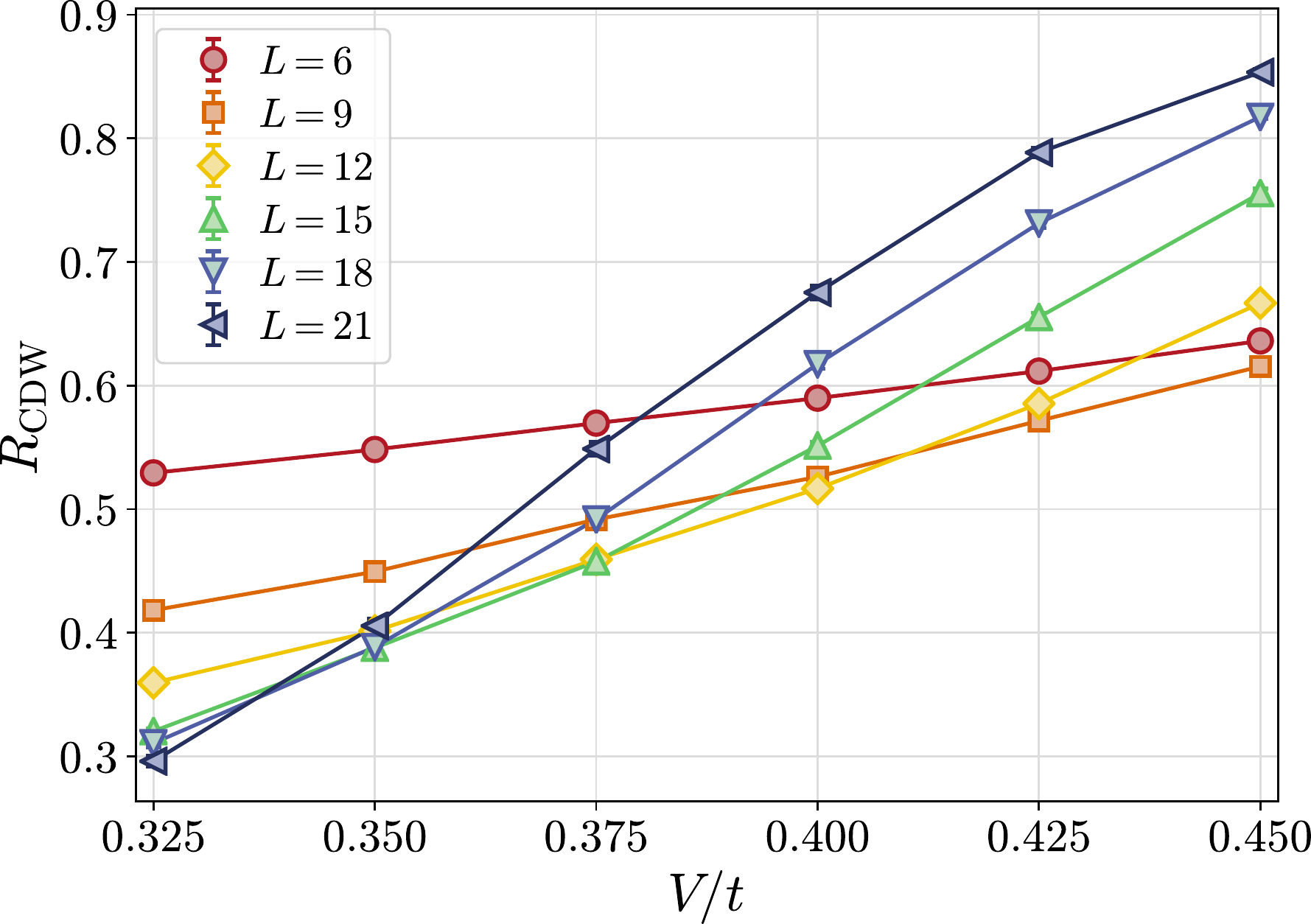}%
	\caption{QMC results for the correlation ratio $R_\mathrm{CDW}$ as a function of $V$ for different systems sizes at ${t''/t = 1/2}$. Error bars are smaller than the symbol size.
	\label{qmc_R}}
\end{figure}

\begin{figure}[t]
	\centering
	\includegraphics[width=0.96\columnwidth]{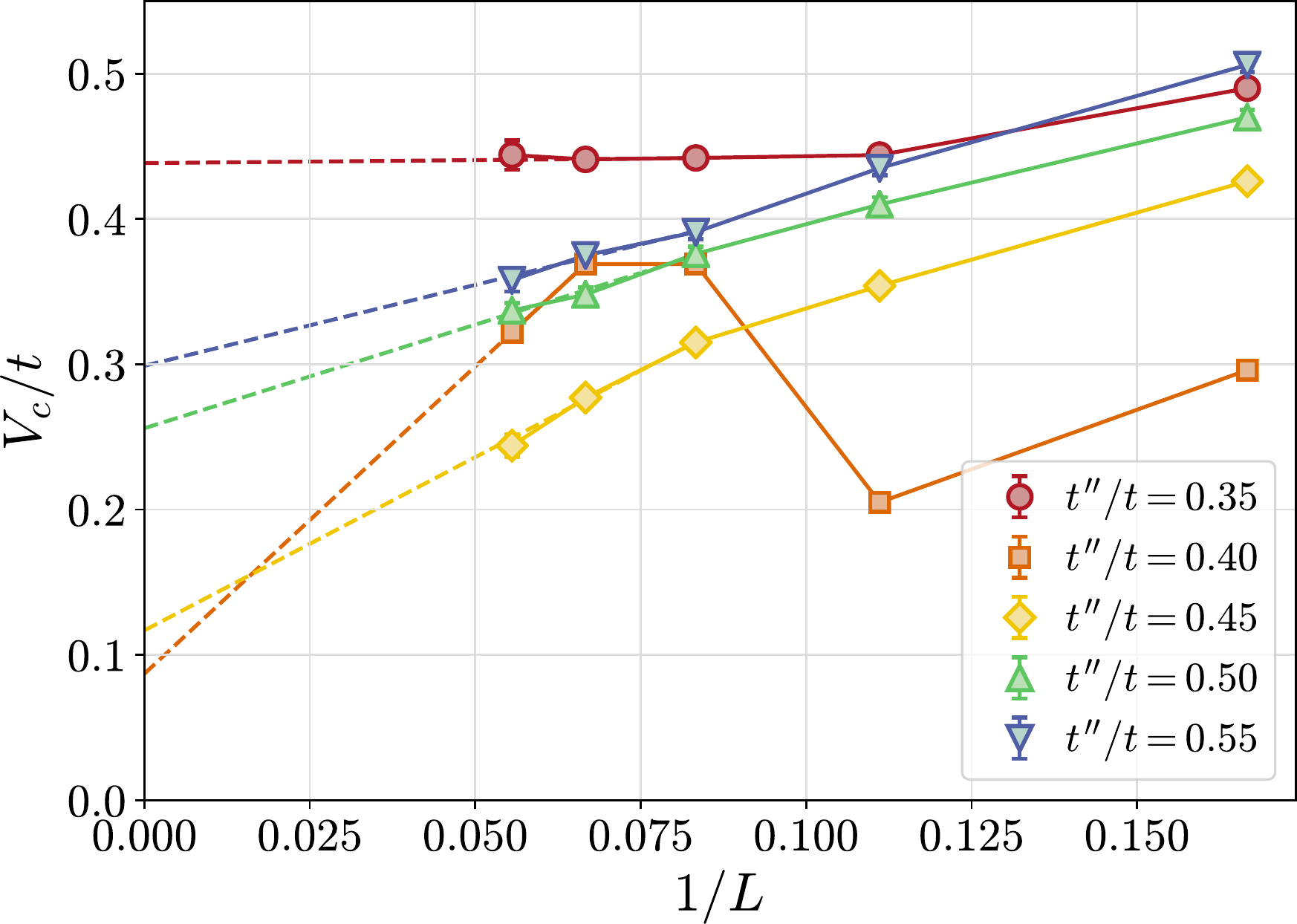}
	\caption{QMC results for the crossing points in the correlation ratio $R_\mathrm{CDW}$ of consecutive system sizes $L$ and ${L+3}$ for various values of ${t''/t}$. Dashed lines indicate linear extrapolations in the low-$1/L$ regime to extract the thermodynamic limit value of the CDW threshold interaction $V_c$.
	\label{qmc_vc}}
\end{figure}

We finally turn to a QMC analysis of the spinless fermion ${t\mbox{-}t''\mbox{-}V}$ model, for which we focus in particular on the identification of the critical interaction strength $V_c$ for the onset of CDW order upon varying the hopping parameter $t''$. For this purpose, we use a projective QMC approach in continuous time~\cite{Wang15,Schuler19} to perform sign-problem-free simulations of the half-filled system on finite honeycomb lattices with ${N = 2 L^2}$ lattice sites and periodic boundary conditions, with linear system sizes up to ${L=21}$. We measure the CDW structure factor
 \begin{equation}\nonumber
   S_\mathrm{CDW}(\vec{q})=\frac{1}{N} \left\langle \sum_{i,j=1}^N \! \epsilon_i \epsilon_j \! \left(\!n_i-\frac{1}{2}\right) \! \left(\! n_j-\frac{1}{2}\right)\!
   \E^{\I\vec{q}\cdot(\vec{r}_i-\vec{r}_j) \! }\!\right\rangle \,,
\end{equation}
where $n_i$ denotes the local density operator on lattice site $i$, $\vec{r}_i$ the spatial position of the unit cell center to which the lattice site $i$ belongs, and ${\epsilon_i = \pm 1}$, depending on which sublattice ($A$ or $B$) the site $i$ belongs to. In the thermodynamic limit, the transition to a CDW ordered state is signaled by a diverging peak in the structure factor at ${\vec{q} = 0}$. In addition, we can exclude the existence of an extended phase with dominant quantum anomalous Hall correlations \cite{Sun09} from tracking the opening of the single particle excitation gap, which we find to coincide with the onset of CDW order (not shown). To extract the threshold value $V_c$ using a finite-size analysis of the QMC data, we follow \cref{Pujari16}, and analyze consecutive crossing points in the correlation ratio 
\begin{equation}
   R_\mathrm{CDW} = 1 - \frac{S_\mathrm{CDW}(\vec{q}_\mathrm{min})}{S_\mathrm{CDW}(0)} \;,
\end{equation}
where $\vec{q}_\mathrm{min}$, with $|\vec{q}_\mathrm{min}| \propto 1/L$, denotes the minimum non-zero lattice momentum next the CDW ordering momentum ${\vec{q} = 0}$. For the non-interacting QBCP case of ${t''/t = 1/2}$, a plot of the QMC results for $R_\mathrm{CDW}$ as a function of $V$ is shown in \cfig{qmc_R}. 
The values of the crossing points in $R_\mathrm{CDW}$ for the considered consecutive system sizes and several values of $t''/t$ are shown in \cfig{qmc_vc}. For sufficiently large system sizes the crossing points eventually converge to the threshold value $V_c$. Such a convergence is indeed seen in \cfig{qmc_vc} for ${t''/t = 0.35}$, for which the available system sizes extend beyond the relevant length scale. For most other values of $t''/t$, the crossing points exhibit a monotonous finite-size behavior, allowing for a linear extrapolation of the low-$1/L$ behavior to obtain a lower bound on the value of $V_c$. The data at ${t''/t = 0.4}$ exhibit more peculiar finite-size effects, so that in this case we restrict the extrapolation to the thermodynamic limit to the largest two system sizes. Along with the estimate for $V_c$ from the crossing point in $R_\mathrm{CDW}$ for the largest simulated systems, we then obtain a conservative error bar for the threshold value of the interaction.

\begin{figure}[t]
	\centering
	\includegraphics[width=0.96\columnwidth]{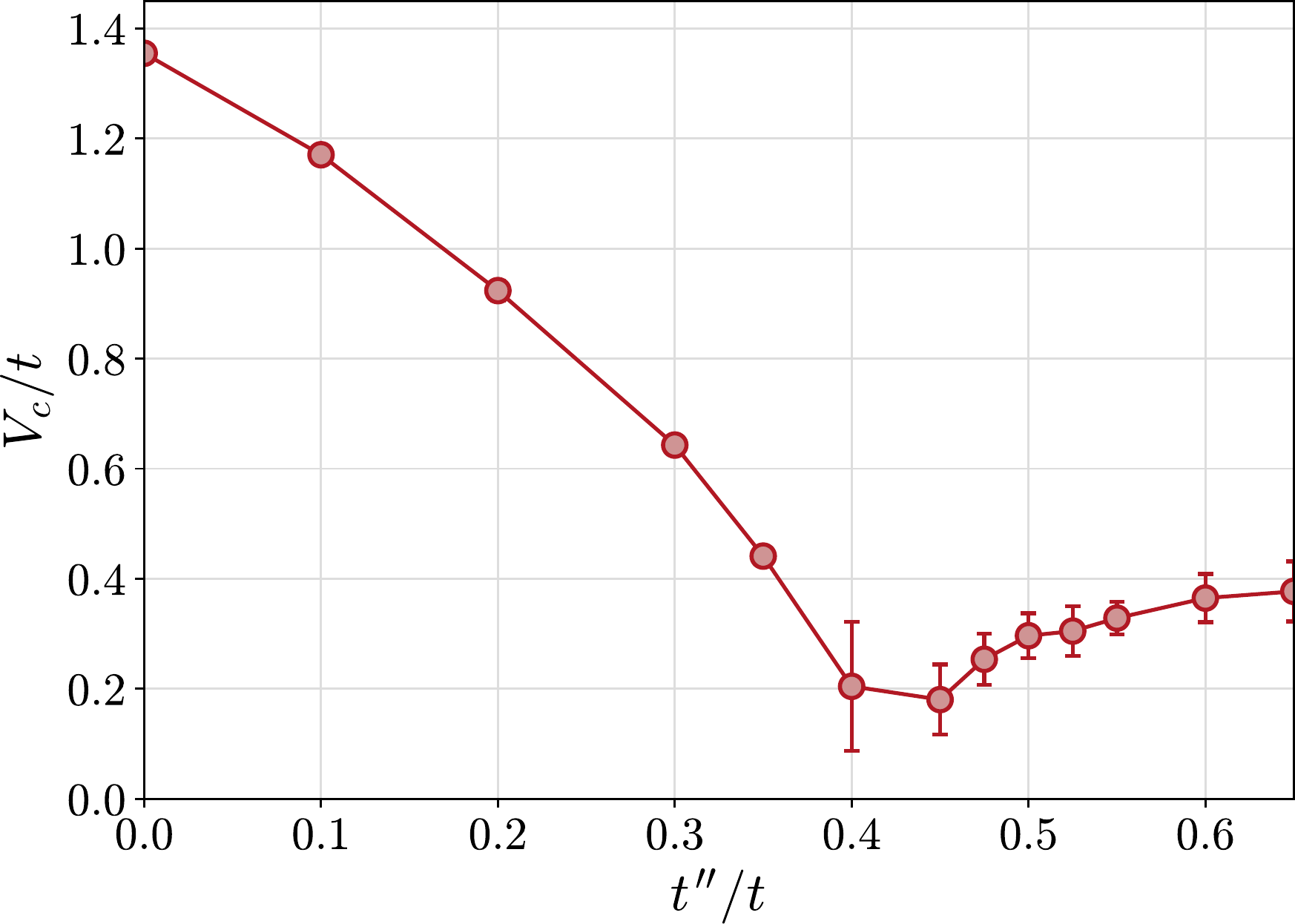}
	\caption{Result for the phase diagram of the ${t\mbox{-}t''\mbox{-}V}$ model from QMC simulations in terms of the extrapolated CDW threshold interaction $V_c$ as a function of $t''$. The lower tick of the uncertainty region (error bar) results from the linear extrapolation, while the upper tick indicates the estimate for $V_c$ from the crossing point in $R_\mathrm{CDW}$ for the largest simulated systems.
	\label{qmc_pd}}
\end{figure}

Based on the QMC data, we then obtain the corresponding phase diagram in \cfig{qmc_pd}, which summarizes our QMC results for $V_c$ as a function of $t''$.
For the case ${t'' = 0}$, our estimate for $V_c$ agrees well with the previously cited result ${V_c/t \approx 1.36}$ from previous QMC studies \cite{Wang14,Hesselmann16}. Note that this value is about a factor of $2$ larger than the RPA estimate. The self-consistent perturbation theory also still underestimates the actual value of $V_c$ by a factor of about $1.5$. For finite values of ${t'' > 0}$, the CDW threshold value $V_c$ is seen to reduce initially. In accordance with the first-order self-consistent perturbation theory, we also extract from the QMC simulations a finite critical value at the non-interacting QBCP condition of ${t''/t = 1/2}$. Compared to the first-order self-consistent perturbation theory, the QMC estimate for ${V_c\approx 0.27t}$ is larger by a factor of about $2$, which is not unexpected in view of a similar underestimation of $V_c$ by the perturbative approach for ${t'' = 0}$. This provides the first evidence from QMC simulations of a finite interaction transition of a system with QBCPs in the free limit, at an energy scale that agrees with estimates from low-order self-consistent perturbation theory.

However, \cfig{qmc_pd} also exhibits a qualitative difference with respect to the first-order self-consistent perturbation theory result: The dip in the $t''$-dependence of $V_c$ in \cfig{qmc_pd} is located to the left of ${t''/t = 1/2}$, at approximately ${t''/t\approx 0.4}$. In perturbation theory, the renormalization of the hopping parameter $t$ in \cfig{Vc1storder} was instead found to lead to a shift of the dip in $V_c$ to the right of ${t''/t = 1/2}$. We trace this difference to restrictions in the available system sizes for the QMC simulations, which limit the ability to resolve the details of the fermion dispersion relation around the split-up QBCPs on the accessible lattice sizes.
\begin{figure}[t]
	\centering
	\includegraphics[width=\columnwidth]{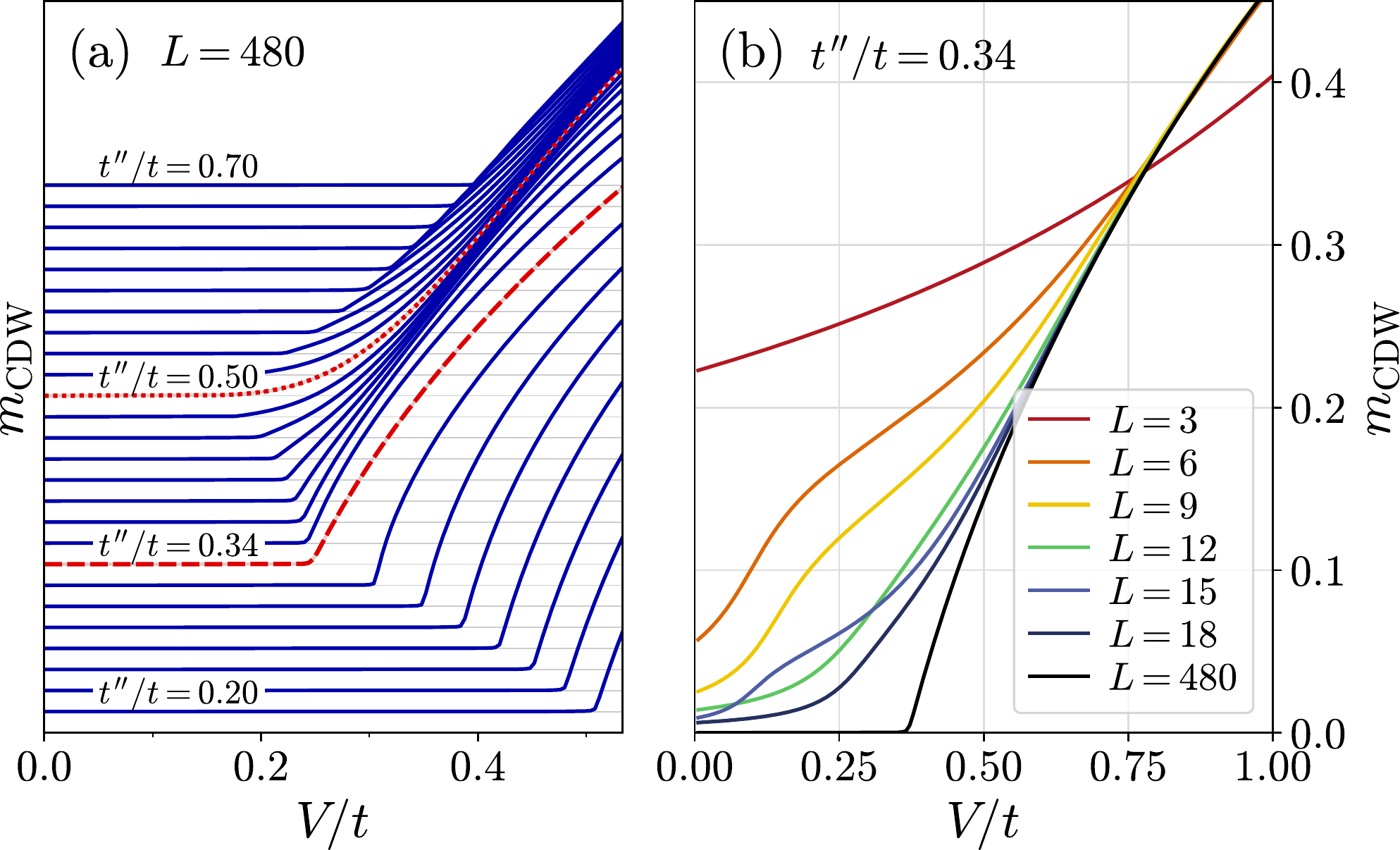}
	\caption{The CDW order parameter $m_\mathrm{CDW}$ as a function of in the interaction $V$ within finite-size Hartree-Fock mean-field theory for (a) different values of detuning ${t''/t}$ and (b) different linear system sizes $L$ in the vicinity of the dip of $V_c$ in \cfig{qmc_pd} at ${t''/t = 0.34}$, which exhibits significant non-monotonous finite size behavior.
	\label{qmc_mft}}
\end{figure}

We can demonstrate such a finite-size effect more explicitly within the Hartree-Fock mean-field approach for the CDW instability on finite systems, which allows us to monitor the finite-size effects over a much wider range of system sizes than accessible to the QMC approach: In \cfig{qmc_mft}(a) we show the normalized CDW order parameter ${m_\mathrm{CDW}=\langle \sum_i \epsilon_i ( n_i-\frac{1}{2}) \rangle/N}$ as a function of the interaction for different values of detuning ${t''/t}$ and a large linear system size ${L=480}$ for which finite size effects are small. The onset of order traces the behavior of $V_c$ in \cfig{chiVcplot}(b). At ${t''/t = 1/2}$ (red, dotted line) the behavior of the order parameter qualitatively changes with respect to other values of ${t''/t}$. The exponential behavior reflects the instability at the perfectly fine-tuned QBCP. 

\begin{figure}[t]
	\centering
	\includegraphics[width=0.94\columnwidth]{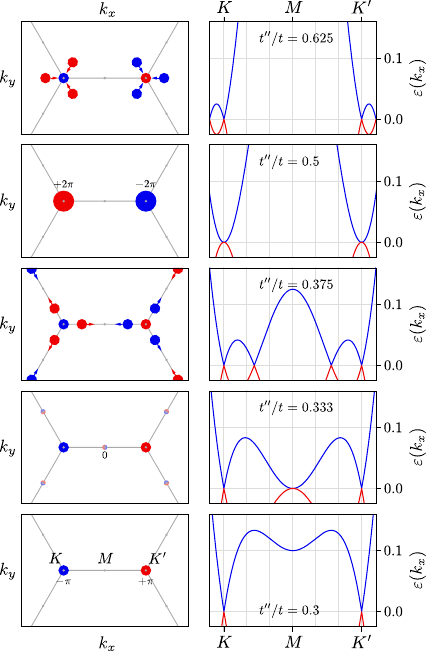}
	\caption{The evolution of the non-interacting dispersion as a function of $t''/t$ shows that the QBCPs are split into four Dirac cones. Arrows indicate the shift of the satellite cones upon decreasing $t''/t$. While one Dirac cone remains at the $K$ ($K'$) point the satellite cones move towards the $M$ point where they eventually annihilate with their counterpart carrying the opposite Berry flux. The size and the color of the circles encode the Berry phase, as indicated in the second panel (from the top) and the bottom panel.
	\label{qmc_berry}}
\end{figure}

In contrast to the results in the previous sections, where optimized momentum resolution has been employed, here we rely on the momenta available on finite size lattices. Figure \ref{qmc_mft}(b) illustrates the effects of this reduced momentum resolution at ${t''/t = 0.34}$ (red, dashed line), i.e., in the vicinity of the dip of $V_c$ in \cfig{qmc_pd}, for the mean-field results. The finite size behavior is distinctively non-monotonous for small system sizes and we anticipate that this also affects the QMC calculations, since the main issue is to resolve the subtle fine-structure of the dispersion relation in the close vicinity of the QBCPs split-up. 

It is also instructive to examine the   evolution of the non-interacting dispersion with $t''$ in terms of the positions of the Dirac cones. 
This is shown in \cfig{qmc_berry}, which  exhibits that the QBCPs, which carry Berry flux ${\pm 2\pi}$, are split into four Dirac cones with Berry flux ${\pm \pi}$ each. While one Dirac cone remains at the $K$ ($K'$) point the three satellite cones move towards the adjacent $M$ points upon reducing $t''/t$,  where they eventually annihilate with their counterparts carrying the opposite Berry flux at $t''/t = 1/3$. This leads to a semi-Dirac dispersion at the $M$ points with zero flux. The annihilation coincides with the kink of $V_c(t'')$ visible in Figs.~\ref{chiVcplot}(b), \ref{Vc1storder} and \ref{qmc_mft}(a). The grid lines in \cfig{qmc_berry}(b) indicate the momentum resolution of an ${L=18}$ lattice and show that the simulations should be able to resolve the Lifshitz energy scale, albeit very sparsely, which can lead to the observed finite size effects and fairly large error bars in \cfig{qmc_pd}. Let us note that while the momentum resolution in the QMC simulations is limited, they do encode the topological structure, i.e., the ${\pm 2\pi}$ Berry flux of the QBCPs in the free single particle Green's function, independently of the lattice size. 

We finally note that a similar shift of the $V_c$-dip away from ${t''/t = 1/2}$ towards ${t''/t \approx 0.4}$ can also result from considering finite temperatures. As shown in \cfig{chiVcplot}(b), in this case, thermal broadening leads to the reduced resolution, i.e., an effective smearing of the details of the dispersion.

\section{Conclusions}

We have investigated the ${t\mbox{-}t''\mbox{-}V}$ as a minimal lattice model with QBCPs, where the interaction introduces first-order Fock self-energy effects that cause a renormalization of the nearest-neighbor hopping. At weak coupling this renormalization induces a Lifshitz transition which splits the QBCPs into multiple Dirac cones. The destruction of the QBCPs cannot be compensated for by fine-tuning of the hopping in order to recover the quadratic band touching in an interacting system. Self-consistent first-order perturbation theory yields an emerging Lifshitz energy scale that is sufficiently large to be properly accounted for in QMC simulations. While subject to finite size effects due to the limited momentum resolution available, this allowed us to identify the critical values for the onset of CDW order from QMC simulations at scales comparable to perturbation theory estimates. The finite size effects, which can be qualitatively reproduced already at the mean-field level, might be overcome by approaches that optimize the momentum resolution of the low energy spectrum \cite{Lang19,Liu19}. This could be assessed in future investigations.

\begin{acknowledgments}
We thank M. Scherer and O. Vafek for discussions. This research was supported by the Austrian Science Fund FWF under the SFB FoQuS (F-4018). Furthermore, we acknowledge support by the Deutsche Forschungsgemeinschaft (DFG) under grant FOR 1807 and RTG 1995, and thank the IT Center at RWTH Aachen University and the JSC J\"ulich for access to computing time through JARA-HPC.

\end{acknowledgments}

\bibliographystyle{apsrev4-1.bst}
\bibliography{tvsigma.bib}

\end{document}